\documentclass[11pt,review]{elsarticle}
\usepackage{ucs} % needed for utf8x
\usepackage[utf8x]{inputenc} 
\usepackage[T1]{fontenc}
\usepackage{lineno,hyperref}
\modulolinenumbers[5]
\journal{arXiv}
\hypersetup{
    colorlinks,
    linkcolor={red!50!black},
    citecolor={blue!50!black},
    urlcolor={blue!80!black}
}

\usepackage[bitstream-charter]{mathdesign}
\DeclareSymbolFont{usualmathcal}{OMS}{cmsy}{m}{n}
\DeclareSymbolFontAlphabet{\mathcal}{usualmathcal}

\usepackage[superscript]{cite}
\usepackage{graphicx,epstopdf,color} %,physics}
\usepackage{amsfonts}
\usepackage{amsmath,mathrsfs}
\usepackage[final,inline]{showlabels} 
\usepackage{mdframed,enumitem}
\usepackage{ifpdf}
\usepackage[normalem]{ulem}
\usepackage{xcolor}
\usepackage{sectsty}
\sectionfont{\normalsize}

\newcommand{\cancel}[1]{\textcolor{red}{#1}}
\renewcommand{\cancel}[1]{}

\RequirePackage[top=20mm,bottom=31mm,left=36mm,right=34mm,head=2mm,includeheadfoot]{geometry}

\begin{document}

\begin{frontmatter}
\setlength{\baselineskip}{15pt}
\title{{\bf Anyons in One Dimension}\\
{\normalsize\rm A Letter in Response to Frank Wilczek's Review Essay in \emph{Inference}}}

\author{Martin Greiter} %\fnref{myfootnote}}
\address{Institute for Theoretical Physics, University of Würzburg, Am Hubland, 
97074 Würzburg, Germany\\[\medskipamount]
{\rm greiter@physik.uni-wuerzburg.de}}

\begin{abstract} 
I give a non-technical account of fractional statistics in one dimension.  In systems with periodic boundary conditions, the crossing of anyons is always uni-directional, and the fractional phase $\theta$ acquired by the anyons gives rise to fractional shifts in the spacings of the relative momenta, ${\Delta p =2\pi\hbar/L\, (|\theta|/\pi+n)}$.  The fractional shift $\theta/\pi$ is a good quantum number of interacting anyons, even though the single particle momenta, and hence the non-negative integers $n$, are generally not.
\end{abstract}

\end{frontmatter}
\setlength{\baselineskip}{15pt}
%\addtolength{\baselineskip}{-6pt}

In Frank Wilczek's marvelous account of anyon physics\cite{wilczek21inference}, he describes how the fascinating possibility for exotic quantum statistics interpolating between the familiar cases of bosons and fermions has evolved from an esoteric exercise in theoretical physics into an experimentally observed concept.  Even though technological applications of the fundamental idea seem far off at present, there is ample prospect that a generalization of the notion, from Abelian to non-Abelian anyons, may play a seminal role in the development of fault tolerant quantum computers\cite{nayak-08rmp1083}.

In this letter, I wish to draw attention to an extension of the concept, from two space dimensions to one.  Just as fractional statistics was considered impossible for more than half a century after the discovery of quantum mechanics, a generalization of the concept to one dimension has long seemed out of reach.  The reason is simply that only in two dimensions, different windings between particles are topologically distinct.  In two dimensions, the amplitudes for paths where identical particles wind clockwise around each other acquire phases opposite to those where the winding is counter-clockwise.  When we exchange particles in one space dimension, however, there is no analog of a winding number.
The only topologically distinct classes of paths, it seems, are those where we interchange the particles, and those where we do not interchange them.  Since interchanging them twice is equivalent to not interchanging them at all, we are left with the familiar cases of bosons and fermions.

There are various manifestations of the fractional phases the wave functions of anyons acquire.  The simplest is fractional relative angular momenta, which in two space dimensions is an even multiple of Planck's quantum $\hbar$ for bosons, an odd multiple for fermions, and an even (or odd) multiple plus a fractional shift for anyons\cite{wilczek82prl957}.  The most striking consequence, however, is the fractional exclusion principle, which for fermions is known as the Pauli principle.  Fermions, as so charmingly explained in Wilczek's review, are anti-social, meaning that each fermion occupies an entire quantum state by itself, while bosons prefer to share the state with as many other identical twins as possible.  For anyons, the exclusion is fractional, and each anyon claims a fraction of a state for itself.

In light of the absence of winding numbers in one dimension, it came as a surprise to many when Duncan Haldane, who subsequently joined Wilczek in the exclusive club of Nobel laureates, discovered in 1991 that the excitations---that is the quasiparticle states above the ground state---of an exactly solvable model of a spin chain exhibit a fractional exclusion principle\cite{haldane91prl937}.  In spin chains, these excitations are called spinons, as they carry the spin of the underlying electrons, but not the charge.  The model in question was discovered independently a few years earlier by Haldane\cite{haldane88prl635} %,haldane-92prl2021} 
and Sriram Shastry\cite{shastry88prl639}.  It is formally extremely appealing, as there is no interaction between the excitations, and all the eigenstates can be identified exactly.  In this sense, the model is as unique to the world of spin chains as the free Fermi sea---the states where the fermions have kinetic energy but interact only through the Pauli exclusion principle---is to electronic states in three dimensional metals. 

For almost two decades, the consensus was that fractional statistics must exist in one dimension, but could only be defined through the generalized Pauli principle introduced by Haldane.  It was not that anyone regarded this situation as particularly satisfying, but no one could even think of a way to resolve the seeming contradiction.  Early this century, another important contribution was made by Robert Laughlin, who had already received a Nobel prize for his theory of the fractionally quantized Hall effect where anyons in two dimensions were discovered recently, and his collaborators\cite{bernevig-01prl3392}.  They obtained explicit wave functions for two-spinon states in the Haldane--Shastry model, and found a very strange interaction term between the spinons.  Interactions, such as the well-known Coulomb interaction between electric charges, usually fall off as some power of the distance.  The interaction term identified by Laughlin and coworkers, however, was proportional to the difference in the momenta of the spinon excitations!

In 2005, Dirk Schuricht, who was a beginning Diploma student at the time, and I, took a closer look at this interaction.  We found that the result by Laughlin and coworkers was in contradiction with a previous result by Fabian Essler, who had shown by a different method that the only interaction between the spinons in this model was statistical \cite{essler95prb13357}.  To be precise, he calculated what is know among experts as the S-matrix using a method applicable to one dimension only, the so-called Bethe Ansatz, and found that it was a constant, the imaginary unit $i$.  Essler's result stated that the spinons were free half-fermions.  This further deepened the mystery around anyons in one dimension.

Not long thereafter, Schuricht and I understood something  significant\cite{%greiter-06prl059701}. %,
greiter-05prb224424}.  
The alleged interaction term between the spinons could be reinterpreted as a fractional shift in the relative momenta of the particles.  To place this into context, consider a single quantum particle with periodic boundary conditions in one dimension, that is, on a ring.  Since the wave function must have a well-defined value everywhere along the ring, the allowed momenta are quantized in units of $2\pi\hbar$ divided by the perimeter.  For the two-spinon states in the Haldane--Shasty model, we found that the momenta of both spinons are shifted towards each other by one quarter of this unit.  These shifts, however, could only result from a very strange
phase the amplitudes would acquire upon interchanges of the spinons.  To be precise, the amplitude would acquire a phase $\pi/2$ if the crossing of spinon 1 occurs counterclockwise relative to spinon 2, 
and $-\pi/2$ if the relative direction of the crossing is reversed.

In other words, we can obtain the shifts in the spinon momenta only if we assign a phase factor with a sign depending on which spinon is number 1 and which is number 2.  This, however, contradicts the principle of indistinguishability, which is so deeply engrained in the principles of quantum mechanics that Wilczek mentions ``The Quantum Mechanics of Identity'' as a potential title he could have used for his essay.  

It took me more than a year to resolve the aparent contradiction\cite{greiter09prb064409}.  The Haldane---Shastry model, and in fact all one dimensional models with anyon excitations, have very special spectra.  Not all the single particle momenta quantum mechanics would permit are available.  In the Haldane--Shasty model, for example, the allowed single spinon momenta amount to only half the values which naive quantization around the ring would permit.  The spectrum, that is the energies of the single particle excitations as a function of momentum, is always either entirely concave or entirely convex in these models.  In the space of available momenta, the group velocity---that is, the velocity a wave packet build out of %these 
excitations would have---is therefore always either a monotonically decreasing (or a monotonically increasing) function of the single particle momenta.  When we view velocities in clockwise direction as positive, one of the spinons is hence always faster than the other one, and the crossings are always uni-directional.
It is only in the light of this uni-directionality that we can assign a statistical phase to the crossings of anyons in one dimension.

Stated differently, when we have anyons in one dimension, the identical, indistinguishable particles become dynamically distinguishable, that is, distinguishable through the quantum states they occupy.  They are willing to pay this price in conceptual complexity in order to exhibit fractional statistics!  As it sometimes happens in science, the ingenuity nature built into a model we had at our disposal was by decades ahead of our imagination.  It may not be of much practical use, but the story told in this letter further underlines how subtle Wilczek's \emph{Quanta of the Third Kind} really are.

%\vspace{50pt}
%\pagebreak
\medskip

\renewcommand{\refname}{\normalsize References}
%\bibliographystyle{../../bib/prsty}
%\bibliographystyle{../../bib/SciPost_bibstyle}
%\bibliography{../../bib/book,../../bib/paper,../../bib/proc,../../bib/unpub,../../bib/htc}

\end{document}